\documentclass[epj]{webofc}
\usepackage[utf8]{inputenc}
\usepackage[varg]{txfonts}   
\usepackage{booktabs}
\usepackage{xcolor}
\definecolor{darkred}{rgb}{0.4,0.0,0.0}
\definecolor{darkgreen}{rgb}{0.0,0.4,0.0}
\definecolor{darkblue}{rgb}{0.0,0.0,0.4}
\usepackage[bookmarks,linktocpage,colorlinks,
    linkcolor = darkred,
    urlcolor  = darkblue,
    citecolor = darkgreen]{hyperref}
\usepackage{subfigure}
\wocname{EPJ Web of Conferences}
\woctitle{Lattice2017}
\begin{document}
%
\selectlanguage{english}
\title{%
Light meson form factors at high $Q^2$ from lattice QCD
}
\author{%
\firstname{Jonna} \lastname{Koponen}\inst{1}\fnsep\thanks{Speaker, \email{jonna.koponen@roma2.infn.it}} \and
\firstname{Andr\'e} \lastname{Zimermmane-Santos}\inst{2} \and
\firstname{Christine} \lastname{Davies}\inst{3} \and
\firstname{G. Peter} \lastname{Lepage}\inst{4} \and
\firstname{Andrew}  \lastname{Lytle}\inst{3}
}
\institute{%
INFN, Sezione di Roma Tor Vergata, Via della Ricerca Scientifica 1, I-00133 Roma, Italy
\and
S\~ao Carlos Institute of Physics, University of S\~ao Paulo, PO Box 369, 13560-970, S\~ao Carlos, SP, Brazil
\and
SUPA, School of Physics and Astronomy, University of Glasgow, Glasgow, G12 8QQ, UK
\and
Laboratory for Elementary-Particle Physics, Cornell University, Ithaca, New York 14853, USA
}
\abstract{%
  Measurements and theoretical calculations of meson form factors are essential for our
  understanding of internal hadron structure and QCD, the dynamics that bind the quarks
  in hadrons. The pion electromagnetic form factor has been measured at small space-like
  momentum transfer $|q^2| < 0.3$~GeV$^2$ by  pion scattering from atomic electrons and
  at values up to $2.5$~GeV$^2$ by scattering electrons from the pion cloud around a proton. 
  On the other hand, in the limit of very large (or infinite) $Q^2=-q^2$, perturbation theory
  is applicable. This leaves a gap in the intermediate
  $Q^2$ where the form factors are not known.

  As a part of their 12 GeV upgrade Jefferson Lab will measure pion and kaon form factors
  in this intermediate region, up to $Q^2$ of $6$~GeV$^2$. This is
  then an ideal opportunity for lattice QCD to make an accurate prediction ahead of the
  experimental results. Lattice QCD provides a from-first-principles approach to calculate
  form factors, and the challenge here is to control the statistical and systematic
  uncertainties as errors grow when going to higher $Q^2$ values.

  Here we report on a calculation that tests the method using an $\eta_s$ meson, a 'heavy
  pion' made of strange quarks, and also present preliminary results for kaon and pion form
  factors. We use the $n_f=2+1+1$ ensembles made by the MILC collaboration and Highly Improved
  Staggered Quarks, which allows us to obtain high statistics. The HISQ action is also designed
  to have small discretisation errors. Using several light quark masses and lattice spacings
  allows us to control the chiral and continuum extrapolation and keep systematic errors
  in check.
}
\maketitle
\section{Introduction}\label{intro}

The electromagnetic form factor of the meson parameterises the deviations from the behaviour
of a point-like particle when hit by a photon. By determining the form factor at different values
of the square of the 4-momentum transfer, $Q^2$, we can test our knowledge of QCD as a function
of $Q^2$. Measurements of $\pi$ and K form factors are key experiments in the new Jefferson Lab
$12$~GeV upgrade (experiments E12-06-101~\cite{JLABE12-06-101} and E12-09-11~\cite{JLABE12-09-011}).
The pion form factor is known experimentally
but with sizeable uncertainties up to $Q^2 < 2.45$~GeV$^2$, and the new experiment will extend
the $Q^2$ range up to $6$~GeV$^2$.

Lattice QCD calculations have been done at small $Q^2$ (see~\cite{Brandt:2013ffb} for a review)
as doing a calculation at small momenta is
easier because of deteriorating signal to noise at large momentum. In~\cite{pionradius} we studied
the pion form factor close to $Q^2=0$ and determined the charge radius of the pion. The goal of
this study is to provide predictions of the form factors at high $Q^2$ ahead of experiments and to
test the applicability of asymptotic perturbative QCD (PQCD). Here we use the $\eta_s$ meson, a pseudoscalar
meson made of strange quarks, as a ``pseudo pion'' to see how (and if) the form factor approaches
the PQCD value. The strange quark is light ($m_s << \Lambda_{\textrm{QCD}}$) from the PQCD point of view, and the behaviour
is expected to be qualitatively similar for $\eta_s$ and $\pi$. The advantage is that strange quarks
are computationally cheaper to simulate on the lattice, and the signal to noise ratio is better
than for lighter quarks. We are now extending the study to pions and kaons, although the maximum
$Q^2$ we can reach in the current calculations is not as high as for the $\eta_s$.

\section{Lattice configurations}

We use lattice ensembles generated by the MILC Collaboration, with 3 different lattice spacings
(ranging from $0.15$~fm to $0.09$~fm) and different light quark masses to allow a reliable
continuum and chiral extrapolation. The Higly Improved Staggered Quark (HISQ) action is used
for both valence and sea quarks, with $u/d$, $s$ and $c$ quarks included in the sea. The strange
quark mass has been tuned to the physical mass by using the $\eta_s$ mass. The ensembles are
listed in Table~\ref{tab:ensembles}.

\begin{table}[thb]
  \small
  \centering
  \caption{Lattice ensembles used in this study: Set 1 is 'very coarse' ($a\sim 0.15$~fm), sets 2
    and 3 'coarse' ($a\sim 0.12$~fm) and set 4 'fine' $a\sim 0.09$~fm. Lattice spacing is set
    using the Wilson flow parameter $w_0=0.1715(9)$~fm. $am_q$ are the sea quark masses in lattice
    units and $L_s/a \times L_t/a$ gives the lattice size in spatial and time directions.
    $M_{\pi}$ and $n_{\textrm{conf}}$  are the pion mass and the number
    of configurations. The last column gives the time extent of the 3-point correlators. More
    details of the lattice ensembles can be found in~\cite{PhysRevD.82.074501,PhysRevD.87.054505}.}
  \label{tab:ensembles}
  \begin{tabular}{cccccccccc}\toprule
 Set & $\beta$ & $w_0/a$  & $am_l$ & $am_s$ & $am_c$ & $L_s/a \times L_t/a$ & $M_{\pi}$ & $n_{\textrm{conf}}$ & $T/a$ \\\midrule
   1 & $5.8$   & $1.1119(10)$  & $0.01300$ & $0.0650$ & $0.838$ & $16\times 48$ & $300$~MeV & $1020$ & $9,12,15$  \\
   2 & $6.0$   & $1.3826(11)$  & $0.01020$ & $0.0509$ & $0.635$ & $24\times 64$ & $300$~MeV & $1053$ & $12,15,18$ \\
   3 & $6.0$   & $1.4029(9)$~~ & $0.00507$ & $0.0507$ & $0.628$ & $32\times 64$ & $220$~MeV & $1000$ & $12,15,18$ \\
   4 & $6.3$   & $1.9006(20)$  & $0.00740$ & $0.0370$ & $0.440$ & $32\times 96$ & $310$~MeV & $1008$ & $15,18,21$ \\\bottomrule
  \end{tabular}
\end{table}

\section{Electromagnetic form factors on the lattice}

\begin{figure}[thb]
  \centering
  \sidecaption
  \includegraphics[width=0.35\textwidth,clip]{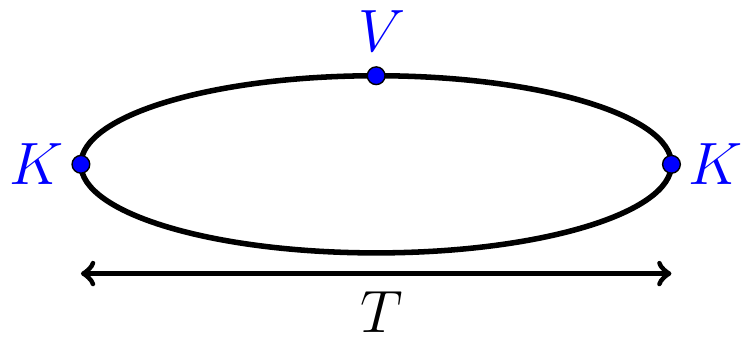}
  \caption{A 3-point correlation function. The meson, here a kaon, is created at
    time $t=t_0$, and destroyed at time $t=t_0+T$. A vector current $V$ is inserted at time $t'$, where
    $t_0 < t' < t_0+T$. We use multiple $T$ values to fully map the oscillating states that are a
    feature of staggered fermions.}
  \label{fig:3ptcorr}
\end{figure}

The electromagnetic form factor is extracted from the 3-point correlation function depicted
in Fig.~\ref{fig:3ptcorr}, where a current $V$ is inserted in one of the meson's quark propagators.
We also need the standard 2-point correlation function of the meson that propagates from time $0$
to time $t$. The 3-point correlation function gives
\begin{equation}
\langle P(p_f)|V_{\mu}|P(p_i)\rangle = F_P(Q^2)\cdot (p_f+p_i)_{\mu},
\end{equation}
where $p_i$ and $p_f$ are the initial and final momenta of the pseudoscalr meson $P$,
respectively. We use a 1-link vector current $V_{\mu}$ in the time direction, and the Breit
frame $\vec{p}_i=-\vec{p}_f$ to maximise $Q^2$ for a given momentum $pa$. This leads to the
simple relation $Q^2=|2\vec{p}_i|^2$. The form factor is normalised by requiring $F_P(0)=1$.

To extract the properties of the meson we use multi-exponential fits with Bayesian priors
to fit both 2-point and 3-point correlators simultaneously. The fit functions are
\begin{align}
C_{\textrm{2pt}}(\vec{p}) &= \sum_i b_i^2 f(E_i(p),t') + \textrm{o.p.t.}, \nonumber \\
C_{\textrm{3pt}}(\vec{p},-\vec{p}) &= \sum_{i,j} \big[b_i(p) f(E_i(p),t) J_{ij}(Q^2) b_j(p) f(E_j(p),T-t)\big]
+ \textrm{o.p.t.}, \nonumber \\
f(E,t) &= e^{-Et}+e^{-E(L_t-t)},
\end{align}
where $E_i$ is the energy of the state $i$ and $\vec{p}$ is the spatial momentum, and o.p.t. stands for
the opposite parity terms. Note that $E_i$ and the amplitudes $b_i$ are common fit parameters for the 2-point
and 3-point functions. We are interested in the ground state parameters $E_o$ (the mass of the meson if
momentum $p=0$), $b_0$ (which is associated with the decay constant of the meson) and $J_{00}(Q^2)$, but use
6 exponentials to make sure that effects of the excited states are properly included in the error estimates.
$J_{00}$ gives the matrix element of the vector current that we need to extract the form factor.
More details can be found in \cite{etasff}.

\begin{figure}[thb]
  \centering
  \sidecaption
  \includegraphics[width=0.42\textwidth,clip]{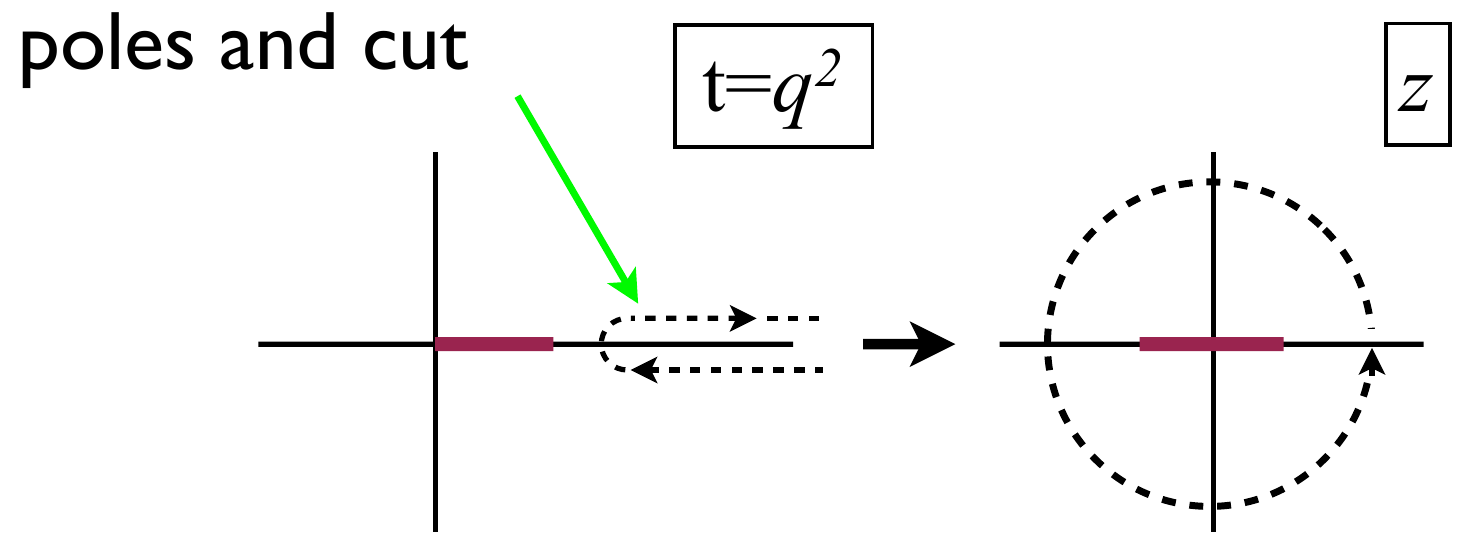}
  \caption{Mapping the domain of analyticity in $t = q^2$ onto the unit circle in $z$.}
  \label{fig:Q2toz}
\end{figure}

To determine the form factor $F$ in the physical continuum limit we must extrapolate in the lattice
spacing and $u/d$ quark mass. We first remove the pole in $F_P(Q^2)$ by multiplying the form factor
by $P_V(Q^2)$, where
\begin{equation}
  P^{-1}_{V}(Q^2)=\frac{1}{1+Q^2/M^2_{V}}.
\label{eq:pole}
\end{equation}
The pole mass $M_{V}$ is the mass of the vector meson that corresponds to the quarks at the current
$V$. If the quarks are light quarks the mass is $M_{\rho}$, if the quarks are strange quarks the pole
mass is $M_{\phi}$. The product $P_VF$ has reduced $Q^2$-dependence because $P^{-1}_{V}$ is a good match
to the form factor at small $Q^2$. We then map the domain of analyticity in $t = q^2$ onto the unit
circle in $z$ --- see Fig.~\ref{fig:Q2toz}:
\begin{equation}
  z(t,t_{\textrm{cut}})=\frac{\sqrt{t_{\textrm{cut}}-t}-\sqrt{t_{\textrm{cut}}}}{\sqrt{t_{\textrm{cut}}-t}
    +\sqrt{t_{\textrm{cut}}}}
\end{equation}
and choose $t_{\textrm{cut}}=4M^2_K$ for the $\eta_s$.
Now $|z| < 1$ and we can do a power series expansion in $z$, and use a fit form
\begin{align}
\label{eq:zfit}
&P_{V}F(z,a,m_{\textrm{sea}})=1+\sum_i z^iA_i\bigg [ 1+B_i(a\Lambda)^2+C_i(a\Lambda)^4+D_i\frac{\delta m}{10} \bigg], \\
& \delta m = \sum_{u,d,s}(m_q-m_q^{\textrm{tuned}})/m_s^{\textrm{tuned}},~\Lambda = 1.0~\textrm{GeV}\nonumber.
\end{align}
The terms with $B_i$ and $C_i$ parametrise lattice discretisation effects and the last
term takes into account possible mistunings of the sea quark masses.

\section{Results}

\begin{figure}[thb] 
  \centering
  \sidecaption
  \includegraphics[width=0.55\textwidth,clip]{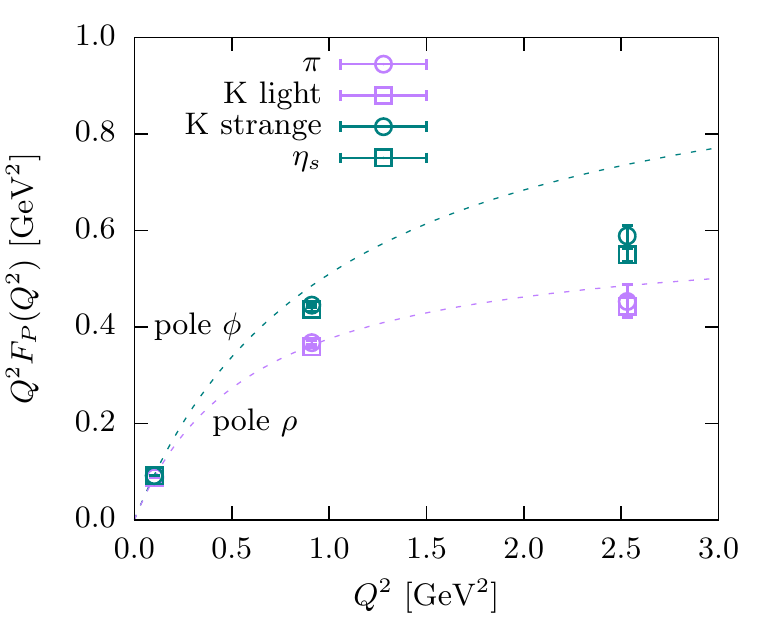}
  \caption{Pion, kaon and $\eta_s$ form factors $Q^2F_P$ on the coarse lattice (set 2). The form factors with
    a strange current are found to be very similar, and so are the form factors with a light current. The
    spectator quark has only very small effect to the form factor. The dashed lines show the corresponding
    pole forms $Q^2P^{-1}_{V}(Q^2)$ (equation~\eqref{eq:pole}) with pole masses $M_{\phi}$ and $M_{\rho}$
    respectively.}
  \label{fig:piKetasFF}
\end{figure}

Figure~\ref{fig:piKetasFF} shows results for pion, kaon and $\eta_s$ form factors $Q^2F_P$.
Let us start by noting how small the effect of the spectator quark is in the pseudoscalar meson
electromagnetic form factor. The pion is made of two light quarks, whereas the $\eta_s$ is made of two
strange quarks. The $K$ meson has one strange quark and one light quark, and the current can thus be
either light or strange. Fig.~\ref{fig:piKetasFF} illustrates how the form factors can be grouped
according to the flavor of the quarks at the current insertion: the $\eta_s$ and the strange-current
$K$ form factors are very similar as are the pion and light-current $K$ form factors. The form factors
follow the pole form at small $Q^2$, but peel away from it when the momentum transfer grows larger.

\begin{figure}[thb] 
  \centering
  \sidecaption
  \includegraphics[width=0.6\textwidth,clip]{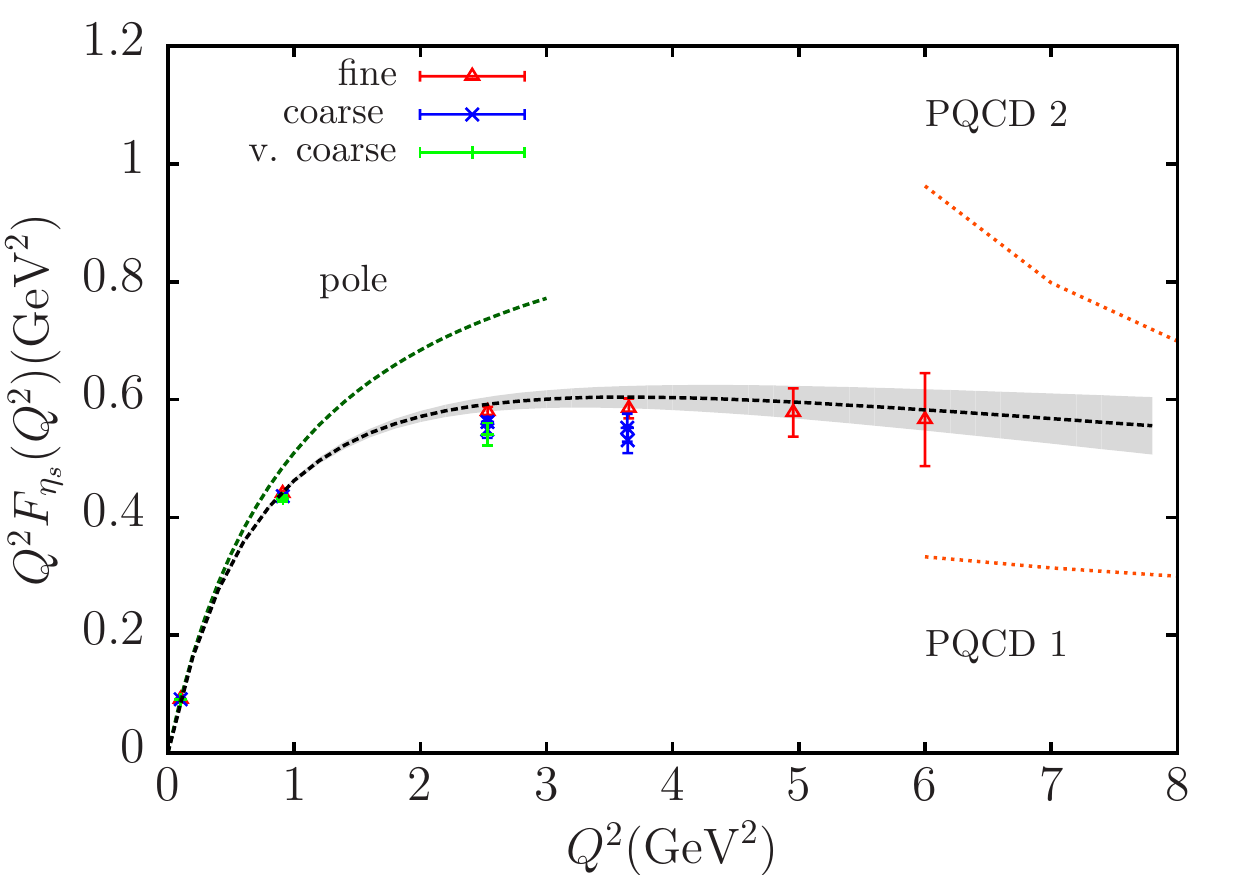}
  \caption{The $\eta_s$ form factor $Q^2F_{\eta_s}$ as a function of $Q^2$. At small $Q^2$ the form factor follows the pole
  form (with pole mass $M_{\phi}$) as expected. The discretisation effects are very small. The grey band
  shows the continuum and chiral extrapolation (equation~\eqref{eq:zfit}). 'PQCD1' is the asymptotic value
  from perturbative QCD, and 'PQCD2' shows the perturbative value with corrections added to the asymptotic
  PQCD. We plot $Q^2F_{\eta_s}$ rather than $F_{\eta_s}$ to compare to the asymptotic value (eq.~\ref{eq:aPQCD}
  multiplied by $Q^2$ gives $8\pi\alpha_s f_P^2$.)}
  \label{fig:etasFF}
\end{figure}

\begin{figure}[thb] 
  \centering
  \sidecaption
  \includegraphics[width=0.55\textwidth,clip]{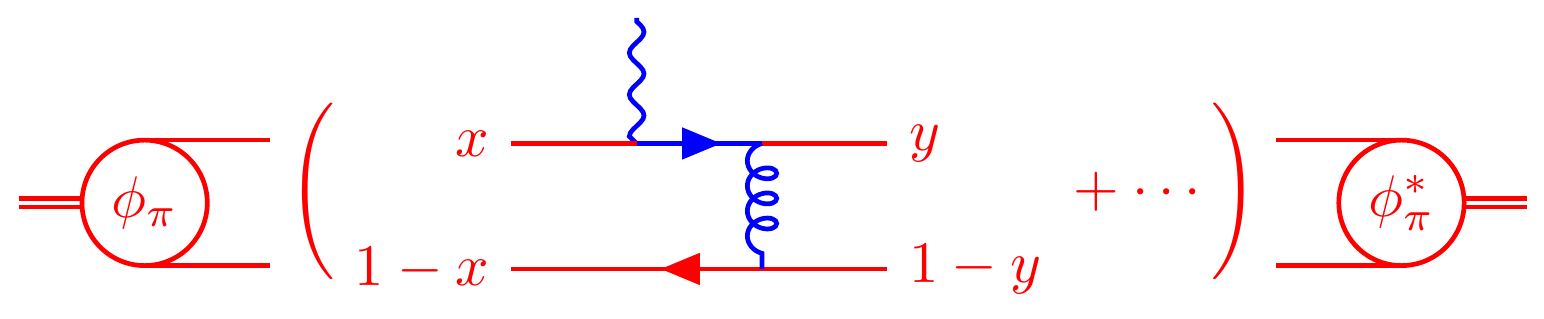}
  \caption{The perturbative QCD description of a meson electromagnetic form factor (here the pion
    is used as an example, but the calculation is analogous for the $\eta_s$). $\phi_{\pi}$ is the
    distribution amplitude and the blue colour marks the high momentum photon and gluon.}
  \label{fig:highmomfact}
\end{figure}

In Figure~\ref{fig:etasFF} we plot the $\eta_s$ form factor obtained on very coarse, coarse and fine
lattices as a function of $Q^2$. We can reach $Q^2\sim 6$~GeV$^2$ on the fine lattice, and the form factor
multiplied by $Q^2$ is found to be almost flat in the $Q^2$ range $3$ -- $6$~GeV$^2$. This can be compared to the
asymptotic value marked with 'PQCD1'. At high $Q^2$ the electromagnetic form factor can be calculated
using perturbative QCD, because the process in which the hard photon scatters from the quark or antiquark
factorises from the distribution amplitudes which describe the quark-antiquark configuration in the meson,
as is illustrated in figure~\ref{fig:highmomfact} using a pion as an example. 
The asymptotic value is
\begin{equation}
  \label{eq:aPQCD}
F_P(Q^2)=\frac{8\pi\alpha_s f_P^2}{Q^2},
\end{equation}
where $f_P$ is the decay constant of the pseudoscalar meson (pion, kaon, $\eta_s$).
The value we obtain for the $\eta_s$ form factor is much higher than the asymptotic value at $Q^2=6$~GeV$^2$.
On the other hand, the curve 'PQCD2' that includes non-asymptotic corrections to the distribution amplitude
lies above the $\eta_s$ form factor. More details can be found in \cite{etasff}.

\begin{figure}[thb] 
  \centering
  \sidecaption
  \includegraphics[width=0.55\textwidth,clip]{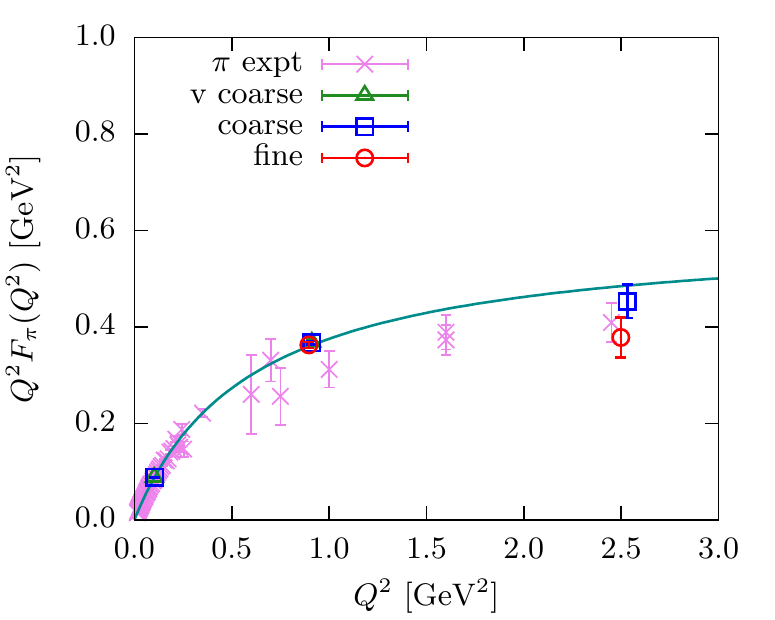}
  \caption{The pion form factor $Q^2F_{\pi}$ as a function of $Q^2$. The agreement with experimental results at
  small $Q^2$ is excellent, and peeling away from the pole form (shown as the continuous line) is observed as expected.
  The results are preliminary as we are pushing to higher $Q^2$, and no continuum extrapolation is done at this time.
  Also smaller light quark masses have to be included in the study to do a reliable chiral extrapolation: the
  pion masses used here are $\sim 300$~MeV. The experimental results are from~\cite{NA7pi,JLAB1,JLAB2}.}
  \label{fig:q2pi}
\end{figure}

\begin{figure}[thb] 
  \centering
  \sidecaption
  \includegraphics[width=0.55\textwidth,clip]{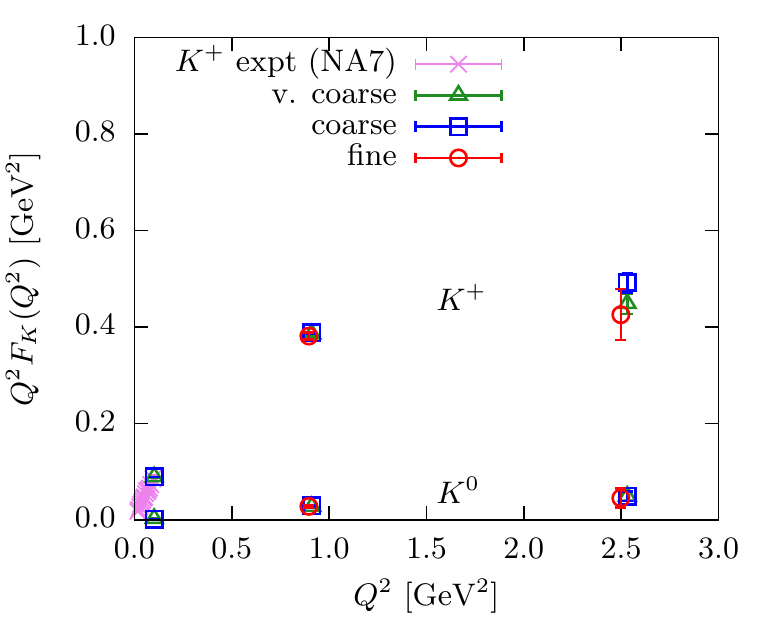}
  \caption{The kaon form factor $Q^2F_K$ as a function of $Q^2$. The agreement with experimental results at small $Q^2$ is
    excellent. The results are preliminary as we are pushing to higher $Q^2$, and no continuum extrapolation is
    done at this time. Also smaller light quark masses have to be included in the study to do a reliable chiral
    extrapolation. The experimental results are from~\cite{NA7K}.}
  \label{fig:q2K}
\end{figure}

In figures~\ref{fig:q2pi} and \ref{fig:q2K} we show our preliminary results for pion and kaon electromagnetic
form factors as a function of $Q^2$. These are the first predictions of the $K^0$ and $K^+$ form factors
from lattice QCD ahead of the Jefferson Lab experiment. The $K^0$ and $K^+$ form factors are calculated from
the strange and light current $K$ form factors by combining with the electric charges of the quarks: $K^+$ is
$u\bar{s}$ and $K^0$ is $d\bar{s}$. Work is underway to go to higher $Q^2$ values and to study the dependence
of the pion and kaon form factors on the light quark mass. The light quark masses used at this preliminary
stage correspond to pion mass of $\sim 310$~MeV. This has been studied in the case of the $\eta_s$ form factor,
where the effect is negligible, but smaller masses are needed to do the chiral extrapolation for the pion and
kaon form factors. We plan to include results from physical light quarks in our final analysis. No continuum or
chiral extrapolation is presented at this time for the pion and kaon form factors.

\section{Conclusions and outlook}

Our $\eta_s$ form factor results indicate that asymptotic perturbative QCD is not applicable at
$Q^2\sim 6$~GeV$^2$ or below --- much larger $Q^2$ are needed. Using strange quarks instead of light quarks
allows us to get some qualitative knowledge of light pseudoscalar meson form factors (pion and kaon form
factors) at high $Q^2$ ahead of the more lengthy calculations required for $K$ and $\pi$. We can also probe higher $Q^2$ values
with strange quarks than with light quarks. However, we can already provide first, preliminary predictions
of the $K^+$ and $K^0$ form factors ahead of the upcoming Jefferson Lab experiment. The pion form factor is the
most challenging. By gathering more statistics and pushing to higher $Q^2$ we will have good theoretical understanding
of the form factors in the momentum range that the Jefferson Lab pion and kaon experiments will use.

\section{Acknowledgements}

We are grateful to the MILC collaboration for the use of their gauge configurations and code.
Our calculations were done on the Darwin Supercomputer as part of STFC’s DiRAC facility jointly
funded by STFC, BIS and the Universities of Cambridge and Glasgow. This work was funded by a
CNPq-Brazil scholarship, the National Science Foundation, the Royal Society, the Science and
Technology Facilities Council and the Wolfson Foundation.



\clearpage
\bibliography{lattice2017}

\end{document}